# High-power liquid-lithium jet target for neutron production

S. Halfon[1,2], A. Arenshtam[1], D. Kijel[1], M. Paul[2,a)], D. Berkovits[1], I. Eliyahu[1],
G. Feinberg[1,2], M. Friedman[2], N. Hazenshprung[1], I. Mardor[1], A. Nagler[1],
G. Shimel[1], M. Tessler[2] and I. Silverman[1]

[1]*Soreq NRC, Yavne, Israel 81800*
[2]*Racah Institute of Physics, Hebrew University, Jerusalem, Israel 91904*

A compact Liquid-Lithium Target (LiLiT) was built and tested with a high-power electron gun at Soreq Nuclear Research Center. The lithium target, to be bombarded by the high-intensity proton beam of the Soreq Applied Research Accelerator Facility (SARAF), will constitute an intense source of neutrons produced by the $^7$Li$(p,n)^7$Be reaction for nuclear astrophysics research and as a pilot setup for accelerator-based Boron Neutron Capture Therapy (BNCT). The liquid-lithium jet target acts both as neutron-producing target and beam dump by removing the beam thermal power (>5 kW, >1 MW/cm$^3$) with fast transport. The target was designed based on a thermal model, accompanied by a detailed calculation of the $^7$Li$(p,n)$ neutron yield, energy distribution and angular distribution. Liquid lithium is circulated through the target loop at ~200°C and generates a stable 1.5 mm-thick film flowing at a velocity up to 7 m/s onto a concave supporting wall. Electron beam irradiation demonstrated that the liquid-lithium target can dissipate electron power areal densities of > 4 kW/cm$^2$ and volume power density of ~ 2 MW/cm$^3$ at a lithium flow of ~4 m/s while maintaining stable temperature and vacuum conditions. The LiLiT setup is presently in online commissioning stage for high-intensity proton beam irradiation (1.91- 2.5 MeV, 1-2 mA) at SARAF.

## I. INTRODUCTION

The $^7$Li$(p,n)^7$Be reaction has been extensively used for the production of epithermal (10-100 keV) neutrons [1,2,3,4,5,6]. For incident proton energies of about 1.91 MeV,

---
[a)] Author to whom correspondence should be addressed. Electronic mail: paul@vms.huji.ac.il

just above the reaction threshold ($E_{thr}(lab)$ = 1.8804 MeV), the thick-target angle-integrated neutron spectrum was shown to have an energy dependence close to $E \cdot exp(-E/E_0)$, peaked at $E_0 \sim$ 25 keV [3,5]. The neutron yield for proton energy in the range 1.9 – 2.0 MeV is of the order of $10^7$-$10^8$ n/s/μA. The similarity of the energy distribution to that of a Maxwell-Boltzmann flux of neutrons at an effective thermal energy $k_BT = E_0 \sim$ 25 keV has important implications for nuclear astrophysics: the $^7$Li(p,n) reaction for proton incident energies of $E_p$= 1.912 MeV is used to mimic a stellar neutron flux typical of that responsible for s-process nucleosynthesis and measure the activation of relevant targets. In a different realm of interest, the near-threshold $^7$Li(p,n)$^7$Be reaction has been widely studied as a prime candidate for production of accelerator-based neutrons for Boron Neutron Capture Therapy (BNCT) [7]. Here the epithermal energy of the emitted neutrons is much closer to that optimally required for therapy of deep-seated tumors ($E_n$= 1 eV - 10 keV) [8,9,10,11] than that obtained from other target materials, such as beryllium ($E_n \sim$ 5 MeV) [8], or from reactor-produced neutrons. An accelerator-based setup for neutron production is also considered more practical for clinical applications than a nuclear-reactor environment.

The low melting point of lithium and its compounds has however been a major drawback in using the $^7$Li(p,n) reaction with high-power accelerators. Use of conventional targets (metallic lithium or compounds such as lithium fluoride) with cooled backing has been usually limited to proton beam intensities of < 100 μA. Blistering of the target backing also sets a limit to high beam-power irradiation of solid lithium targets. With the availability of higher beam intensities (in the range of milliamps) from modern superconducting linear accelerators, the development of lithium targets capable of sustaining high beam powers has become necessary [12,13].

In this paper we describe a Liquid-Lithium Target (LiLiT), designed to dissipate beam powers of several kW for high-intensity neutron production by the $^7$Li(p,n) reaction. The setup (fig. 1) is based on a high-velocity windowless liquid-lithium jet, flowing transverse (vertically) to the incident particle beam (horizontal). The absorbed beam power, conveyed by the lithium flow, is heat-exchanged in a reservoir. The target has been constructed and was tested with a high-power electron beam and it is presently being installed at the SARAF superconducting linear accelerator facility at Soreq NRC, Israel [14] online commissioning.

Section II presents the physical principle of the target, including estimates of the power and power density dissipated by a typical high-intensity proton beam in a thick lithium target and of the temperature distribution profile expected in a liquid-lithium jet flowing at high velocity. Section III describes the technical design and major components. The design was inspired by a liquid-lithium target proposed for fragmentation of heavy-ion beams in future high-power accelerators for radioactive-ion beams [15] and as a fast neutron source for fusion reactor material testing [16] through the IFMIF collaboration [17,18]. Section IV shows results of circulation tests of the liquid lithium and describes power dissipation tests performed with an electron beam, delivering power and power density similar to those expected for a high-intensity proton beam. Section V includes the calculation of the expected neutron spectrum and intensity for the LiLiT setup, when irradiated by the high intensity proton beam at SARAF.

## II. PHYSICAL PRINCIPLE AND THEORETICAL ESTIMATES

The physical principle of the LiLiT system, schematically illustrated in fig. 1, consists of a film of liquid lithium (at ~200 ºC, above the lithium melting temperature of 180.5 ºC) forced-flown at high velocity onto a concave-curvature thin stainless-steel wall. The target is to be bombarded by a high-intensity proton beam impinging directly on the Li-vacuum interface (windowless) at an energy above and close to the $^7$Li($p,n$) reaction threshold. A rectangular-shaped nozzle just before the curved wall determines the film width and thickness (18 mm and 1.5 mm respectively, see Section III for details). The first few μm's at the surface of the liquid-lithium film serve thus as a neutron-producing thick target and the deeper layers as a beam dump from which the power is transported by the flow to a heat exchanger. The setup takes into advantage the exceptionally high specific heat capacity of liquid lithium ($C_p$= 4350 J/kg·K, as high as that of water) and its extremely low vapor pressure ($7\times10^{-9}$ mbar at 220 ºC). A simple expression of the power conveyed is given by $q_{conv} = \rho v C_p A \Delta T$, where $\rho$ is the density of liquid lithium (0.51 g/cm$^3$ at 220 ºC), $v$ its flow velocity, $A$ the cross section area of the film and $\Delta T$ the permitted temperature elevation. For values of $\Delta T$ ~ 100 ºC and $v$ of the order of 1 m/s, the power conveyed by the flow is ~ 6 kW for the jet dimensions given above. The flow velocity is maintained in fact so that the steady-state local temperature in the beam spot area does not exceed a limit value determined by an allowed evaporation rate. We describe in this section the physical processes involved and derive a detailed thermal model of the system.

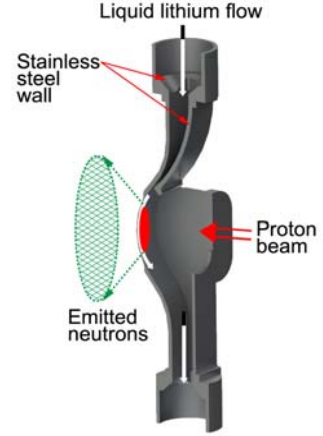

Figure 1: Schematic illustration of the operation of the LiLiT system (see text). For clarity, one side wall of the nozzle region was removed in the drawing. The incident proton beam impinges directly on the liquid lithium flow (windowless) and within the first μm's produces a flux of neutrons, kinematically forward-focused if the incident energy is close to the neutron threshold. The protons are stopped within the lithium film and the beam power (of the order of MW/cm$^3$, see Fig. 4) is conveyed by the flow to a heat exchanger.

## A. The thick target $^7$Li($p,n$)$^7$Be reaction

For proton energies just above the $^7$Li($p,n$) reaction threshold ($Q$= -1.6442 MeV, $E_{thr}(lab)$ = 1.8804 MeV), the produced thick-target neutrons are emitted in the forward direction, with the most probable neutron emission angle between 20° to 30° and most probable energy between 25 and 30 keV [3]. Fig. 2 shows the maximum and median neutron emission angle together with the total neutron yield, calculated from the reaction kinematics as a function of proton energy up to 1.92 MeV. According to Fig. 2, at 1.91 MeV, the maximum and median neutron emission angles are 60.2° and 27.6°, respectively, with a total yield of 2.4 × 10$^{10}$ n/mC [6]. A detailed description of the neutrons angular and energy distribution yield for 1.91 MeV protons bombarding a thick lithium target is shown in Fig. 3. The near-threshold yield is calculated following the method described by Lee and Zhou [3] (see also [6]).

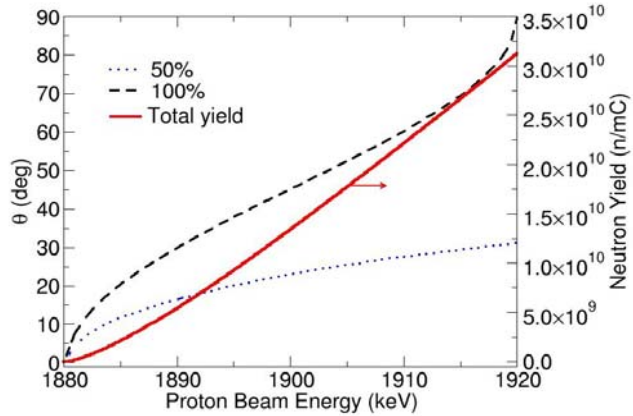

Figure 2: The maximum (dashed) and median (dotted) neutron emission angle from the $^7$Li(p,n) reaction and the neutron yield (solid) as a function of proton energy above threshold ($E_{th}$= 1.8804 MeV).

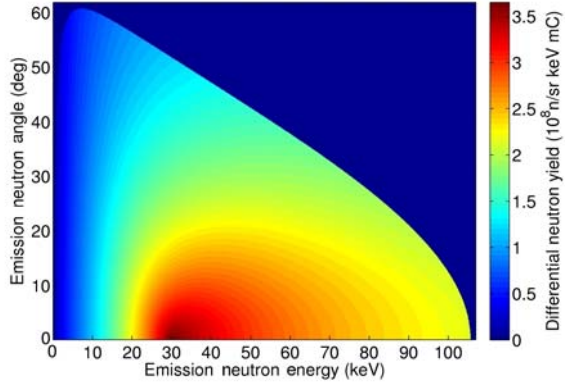

Figure 3: Double differential neutron yield ($d^2N/d\Omega dE$) per mC for 1.91 MeV protons bombarding a thick lithium target. The abscissa is the neutron emission energy and the left ordinate is the neutron emission angle. The discontinuity between zero and non-zero yield is due to the gridded calculation method [3] and the assumed discrete proton energy.

## B. Proton energy loss in lithium

Fig. 4 illustrates the differential energy loss for a proton beam as a function of depth in liquid lithium, calculated using the Monte Carlo code TRIM [19], and the corresponding power density profile. Nominal values of the incident energy and intensity of the proton beam are taken in this paper as 1.91 MeV and 1 mA, respectively (Table I). The plotted quantity on the left axis of Fig. 4, *i.e.* the mean differential energy loss for a beam of incident 1.91 MeV protons on a thick lithium target, is different from the true value (*dE/dx*), especially near the end of the particle range, because of significant effects of energy and angle straggling. The maximum of the mean energy loss in Fig. 4 is 33 keV/μm while the maximum of the differential energy loss (*dE/dx*), on which the TRIM calculation is based, is 48.2 keV/μm. Neutrons are generated by the $^7$Li(*p,n*)$^7$Be reaction within ~ 4 μm of the incident surface, after which the proton energy is below the $^7$Li(*p,n*) threshold. The proton range in liquid lithium until stopping is 154 μm, within which the bulk of the beam power is absorbed, with a significant fraction in a thin layer around the Bragg peak. For a 1 mA beam having a transverse radial Gaussian profile with a standard deviation $\sigma_r$ = 2.8 mm (Table I)), the peak volume power density averaged within one standard transverse deviation $\sigma_r$ at the Bragg peak is ~0.67 MW/cm$^3$. The two-dimensional profile of the power deposition, calculated in these conditions, is illustrated in Fig. 5. Note that, due to the much smaller transverse beam size and the narrower Bragg peak, the volume power density required here is larger than the value planned (< 0.3 MW/cm$^3$) in the

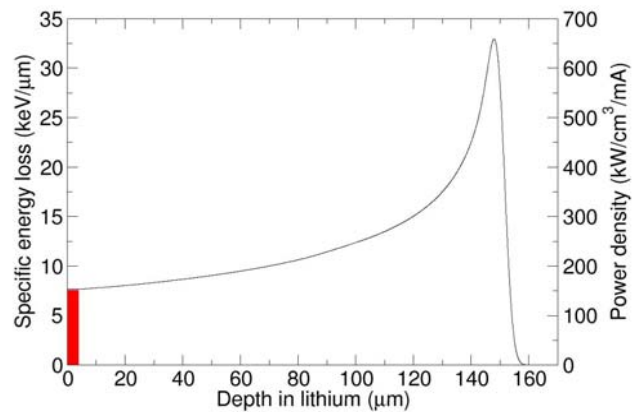

Figure 4: Mean specific energy loss incurred by a beam of 1.91 MeV protons at normal incidence as a function of depth in liquid lithium (density = 0.51 g/cm$^3$). The shallow band represents the region (~4 μm deep) of neutron production by the $^7$Li(*p,n*) reaction above the reaction threshold of 1.8804 MeV. The right-hand scale represents the volume power density deposited in liquid lithium as a function of depth by a proton beam of 1 mA with a transverse radial normal distribution having a standard deviation $\sigma_r$ = 2.8 mm. The volume power density plotted is the mean power deposited within one radial standard deviation $\sigma_r$.

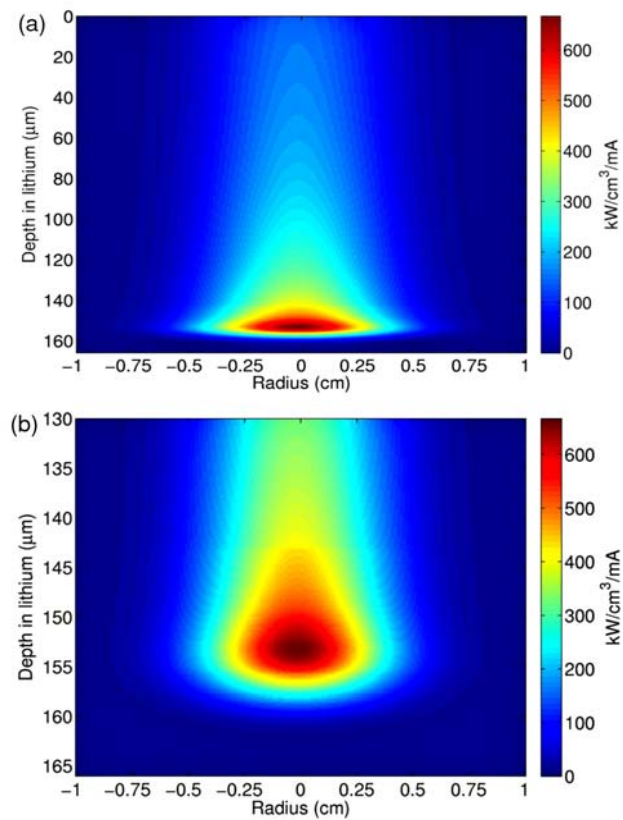

Figure 5: (a) Two-dimensional (longitudinal vs. transverse direction) density plot of the volume power density deposited by a 1 mA proton beam at incident energy of 1.91 MeV on a liquid-lithium target with a radial intensity Gaussian distribution ($\sigma$= 2.8 mm); (b) enlarged plot of the Bragg peak region.

IFMIF design [20] with a 10 MW, 40 MeV deuteron beam and a flattened footprint of 100 cm$^2$.

TABLE I: Nominal specifications of proton beam used in this work.

| | |
|---|---|
| Energy | 1.91 MeV |
| Intensity | 1 mA |
| Power | 1.91 kW |
| Beam Size (radial standard deviation) | 2.8 mm |
| Max power density in lithium target | 0.67 MW/cm$^3$ |

### C. Thermal model of liquid lithium free surface target

A thermal model was built to estimate the flow velocity required to prevent excessive heating and evaporation of lithium in the beam spot area and possible upstream diffusion of vapors in the accelerator beam line. Two limits are relevant when the thermal performance of a liquid target is estimated: the boiling temperature that depends on the fluid pressure and the evaporation rate of the fluid, a phenomenon that increases exponentially with the surface temperature.

Two types of liquid boiling are considered: heterogeneous boiling, where gas cavities on solid surfaces in contact with the liquid lithium start to burst and homogeneous boiling, where spontaneous clusters of high energy lithium atoms form vapor bubbles in the bulk liquid. Momozaki et al. [21] studied the conditions for homogeneous boiling in heated windowless lithium free jets and concluded that it does not occur for liquid temperature below 625°C, which is several hundred degrees higher than the saturation temperature at normal operating conditions of 350°C at ~2 × 10$^{-5}$ mbar. With regard to heterogeneous boiling in confined jets such as the LiLiT setup, the lithium temperature at the external walls is low (~350°C) and boiling is not expected to initiate there. Heterogeneous boiling could however initiate on particles formed by contaminants in the liquid though these are usually quite rare in clean liquid metals.

The calculation of the liquid lithium surface temperature, needed to estimate the evaporation rate, as a function of beam parameters and flow velocity requires solution of the energy conservation equations for the lithium jet expressed by the equation:

$$\rho C_p (\frac{\partial T}{\partial t} + \underline{v} \cdot \nabla T) = \nabla \cdot (\lambda \nabla T) + q \quad , \tag{1}$$

where $\lambda$ is the heat conductivity of liquid lithium and $q$ the heat source per unit volume due to the particle beam. Several approximations are made to simplify the solution of Eq. 1. The very low Prandtl number of liquid metals (of the order of 0.01) determines that heat transfer across the jet (perpendicular to the main flow direction) is mainly due to conduction rather than convection. Hence, local velocity distributions due to turbulence and boundary layers on solid walls have negligible influence on heat transfer and temperature distribution across the jet. We can also neglect friction on the free surface of the liquid lithium (vacuum interface) or with the walls (out of the heated zone) and assume a plug flow model with constant velocity across the jet for heat conduction estimations. It is understood that the boundary layers on the walls that guide the flow will contribute to momentum exchange and slow down of the flow but this effect can be solved separately. Moreover, the heat convection term $q_{conv} = \rho v C_p \Delta T$, estimated above as ~ 6 kW for a velocity of 1 m/s is much larger than the conduction term in the flow direction, $q_{cond} = \lambda \cdot \partial T / \partial z$ ~ 10 W, in the same conditions for a temperature gradient of 100 K/cm. In summary, for high flow velocity, heat conduction in the direction of the flow can be neglected compared to convection while for any other direction, convection can be neglected relative to the conduction term. Given these approximations, Eq. 1 reduces to the following single partial differential equation:

$$v(x) \rho C_p \frac{\partial T}{\partial x} = \frac{\partial}{\partial y}(\lambda \frac{\partial T}{\partial y}) + \frac{\partial}{\partial z}(\lambda \frac{\partial T}{\partial z}) + \frac{\partial E}{\partial z} I_0 e^{-(r^2/2\sigma^2)} \quad , \tag{2}$$

where $x$ is the jet flow direction, $v(x)$ its local velocity, $y$ the transverse distance and $z$ the particle beam direction. The third term on the right-hand side is the heat source per unit volume due to a radial Gaussian beam load, ($r = \sqrt{(x-x_0)^2 + (y-y_0)^2}$), $I_0$ the maximum particle flux density at the beam center and $\sigma$ the radial standard deviation of the distribution. The effect of heat loss by radiation from the front side of the jet on the energy balance is accounted for by the boundary conditions and the same applies to energy loss due to lithium evaporation. Heat loss by radiation or conduction to walls on all other sides of the jet can be neglected based on the expected low temperature and gradients in these zones. The proton specific energy loss in lithium ($\partial E / \partial z$) was calculated as described in Sec. II C, using the code SRIM [19]. The lithium initial temperature was taken as 220°C and the flow velocity as 4 m/s. Temperature profiles of the lithium target calculated using equation (2) for a beam power of 1.91 kW and beam transverse width σ = 2.8 mm, are shown in Fig. 6. The peak temperature (around 320 °C) is located near the surface of the liquid lithium, ~ 3 mm downstream from the center of the beam. The hottest area (315-320 °C) is a transverse ellipse near to the irradiated surface of the lithium with a major axis of ~5 mm, a minor axis of ~2 mm and a depth of 0.1 mm. These distribution calculations were repeated for various lithium flow velocities and fig. 7 shows the dependence of the peak temperature on flow velocity: in these conditions, a flow velocity above ~ 3 m/s is required to ensure that the maximum temperature remain below (homogeneous) boiling ($T$= 350 °C at a typical working pressure of $2 \times 10^{-5}$ mbar [22]).

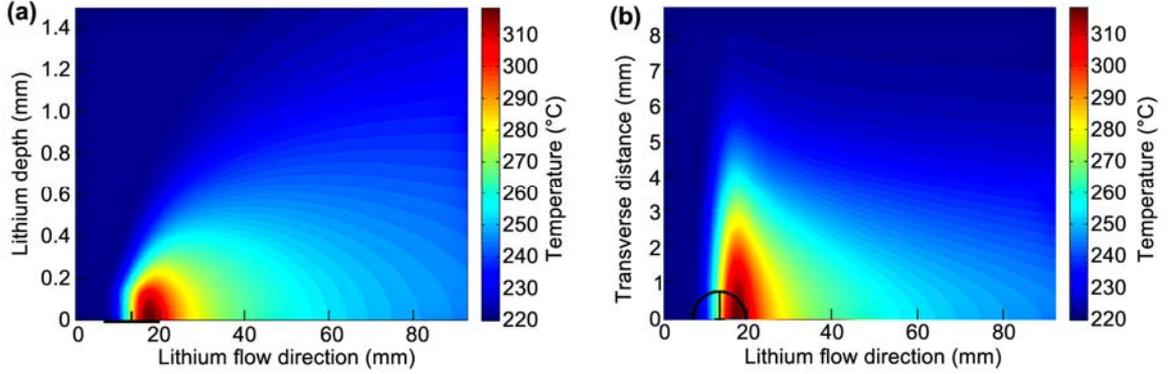

Figure 6: Temperature profiles in the liquid lithium target irradiated by a 1 mA proton beam at incident energy of 1.91 MeV with a transverse intensity radial Gaussian distribution (σ = 2.8 mm), calculated using Eq. (2) for a flow velocity $v$ = 4 m/s. The horizontal axis is the lithium flow direction ($x$ in Eq. (2)). The vertical axis is: (a) the depth in lithium ($z$ in Eq. (2)); (b) the transverse distance ($y$ in Eq. (2)) in (b). The full beam spot area (>95% of beam intensity), 11 mm in diameter and centered at $x$ = 13.5 mm, is marked in (a) by a solid line below the horizontal axis, and the center of the beam is marked by a perpendicular thin solid line. In (b) the beam spot is marked on the figure as a half solid circle.

The effect of evaporation is not considered significant regarding the energy balance but as mentioned above, it must be calculated to assess operational risks. The maximum atom flux evaporating from a liquid (*i.e.* in ideal vacuum and assuming no condensation on the liquid) is expressed by the Hertz-Knudsen equation (see for example [23,24]):

$$\frac{\partial \dot{n}}{\partial S} = \frac{P_v}{\sqrt{2\pi m k_B T}} \quad \text{(atoms/m}^2\cdot\text{s)}, \tag{3}$$

where $P_v$ is the vapor pressure of the liquid and $m$ is the mass of an atom. The mass evaporation rate, calculated by integrating Eq. (3) over the temperature distribution on the lithium surface (fig. 6b) for different flow velocities and using the temperature dependence of $P_v$ (see [22]), is also illustrated in fig. 7. For flow velocities above ~ 3 m/s ($T_{max}$ ~ 350 °C), the estimated evaporation rate is less than 0.1 mg/h, expected to allow safe operation for long periods. Notwithstanding these results, we included a vapor trap (described in the next section) between the target vacuum chamber and the accelerator beam line in order to reduce migration of lithium vapor towards the accelerator.

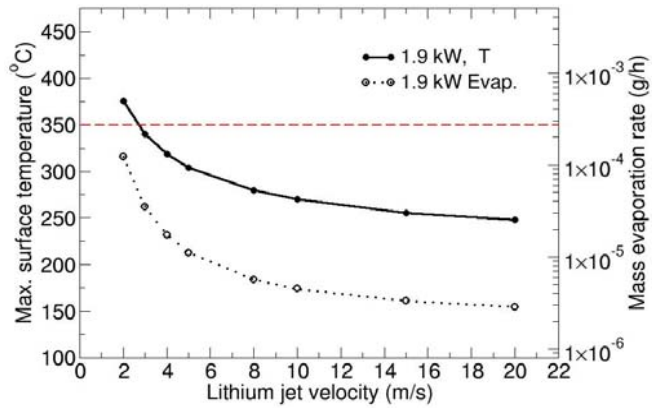

Figure 7: Lithium peak surface temperatures (solid dots and solid line) and mass evaporation rate (open dots and dotted line) as a function of jet velocity for nominal operating conditions (1.91 kW, $\sigma$ = 2.8 mm), calculated respectively using Eqs. (2) and (3). The mass evaporation flux (in g/cm$^2$.h) is extracted from (3) using the numerical expression $d\dot{m}/dS = 1.58\sqrt{A/T} \cdot P_v$, where A is the atomic mass of lithium in grams (6.94 g), $T$ in K and $P_v$ in Pa. The dashed horizontal line corresponds to the boiling point temperature of ~350°C at a typical operating pressure of ~2×10$^{-5}$ Torr.

## III. LiLiT: DESIGN AND DESCRIPTION

Fig. 8 illustrates the LiLiT general assembly, consisting of a loop (G) of circulating liquid lithium maintained above liquefaction temperature (188 °C) by external and internal heating elements. The liquid lithium flow is driven by an electromagnetic (EM) induction pump (F) from the reservoir (E) through the pipes (~2.54 cm in diameter) and the vacuum chamber (B) that hosts the nozzle (D). The lithium is returned into the containment tank where a heat exchanger removes the beam power to a secondary oil loop. The vacuum in the system is maintained around 10$^{-5}$-10$^{-6}$ mbar using a custom-built arc-pump and an adjoined ion-pump. Most parts of the system were fabricated from stainless steel 316 and vacuum gaskets made of soft iron, both compatible with liquid lithium [25]. We describe below the system starting from the lithium reservoir and following the lithium flow (Sec. A-E).

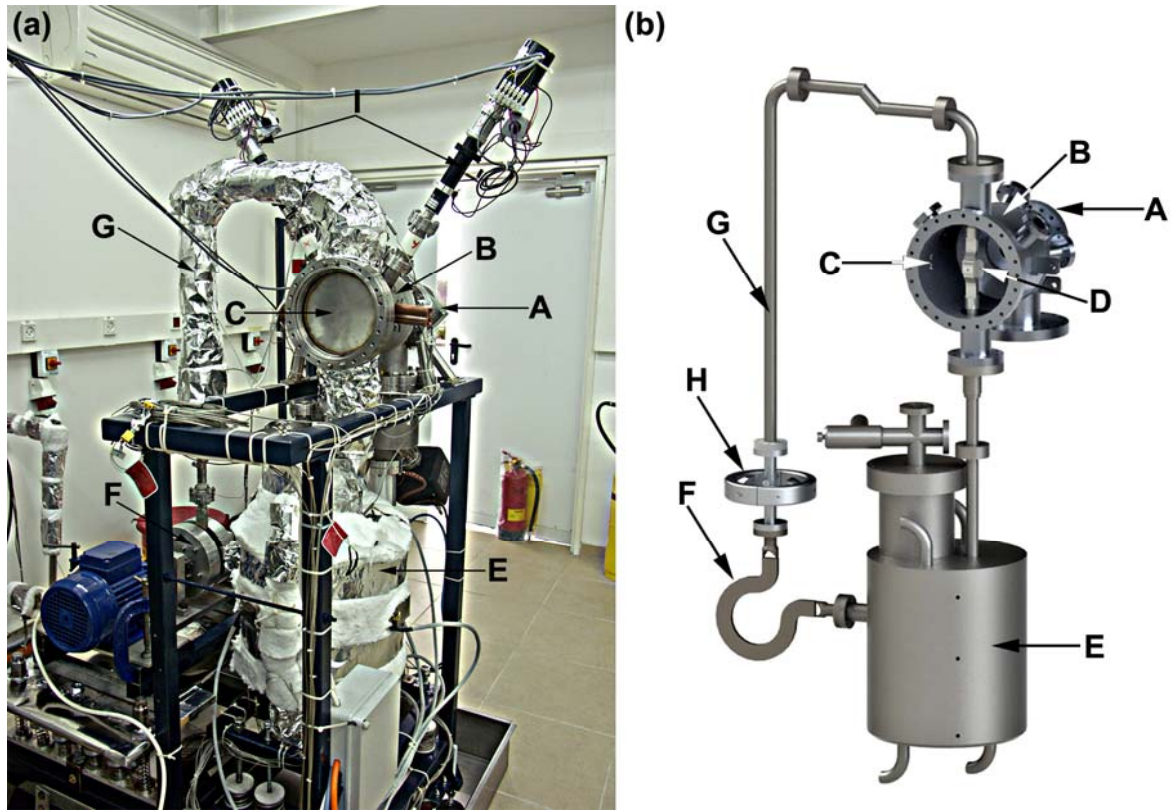

Figure 8: Photograph (a) and schematic drawing (b) of the LiLiT assembly viewed from the neutron exit port: (A) - proton beam inlet port; (B) - target chamber; (C) - neutron port (closed in (a) and open in (b)); (D) - lithium nozzle, positioned in the chamber near the inner side of the exit port (not seen in (a)), (E) - lithium containment tank (including heat exchanger and $^7$Be cold trap); (F) - Electromagnetic (EM) pump (only the circulation loop is shown in (b)); (G) - loop line; (H) - electromagnetic flow-meter (not shown in (a)); I- drivers for beam diagnostics- a wire scanner and a Ta diagnostic plate (not seen in (b)).

### A. Lithium reservoir

The lithium metal (~7.5 kg, 15 liters) is loaded in a 20-liters cylindrical reservoir (Fig. 9). The reservoir cylinder is covered by three ceramic heaters with total heating power of 7.5 kW. The heat exchanger, located in the upper part of the reservoir, is shaped as a cylindrical sleeve, circulating separately a synthetic oil coolant [26] and surrounding the reservoir (see flow diagram in fig. 9b). A cold trap (Fig. 9(b)), designed to trap radioactive $^7$Be, is a cylindrical tank, placed in the lower part of the reservoir. Lithium flows in tubes crossing the cold trap tank while the mineral oil separately circulates around them (fig. 9b). The oil continues up to the heat exchanger and then flows out of the reservoir toward an oil-air heat exchanger (fan) designed for removal of up to 10 kW heat power. The temperature in the cold trap is designed to be the lowest in the lithium loop (around 190°C). Radioactive $^7$Be is expected to accumulate mostly in the coldest area of the loop (see Sec. III F). This effect will be studied experimentally.

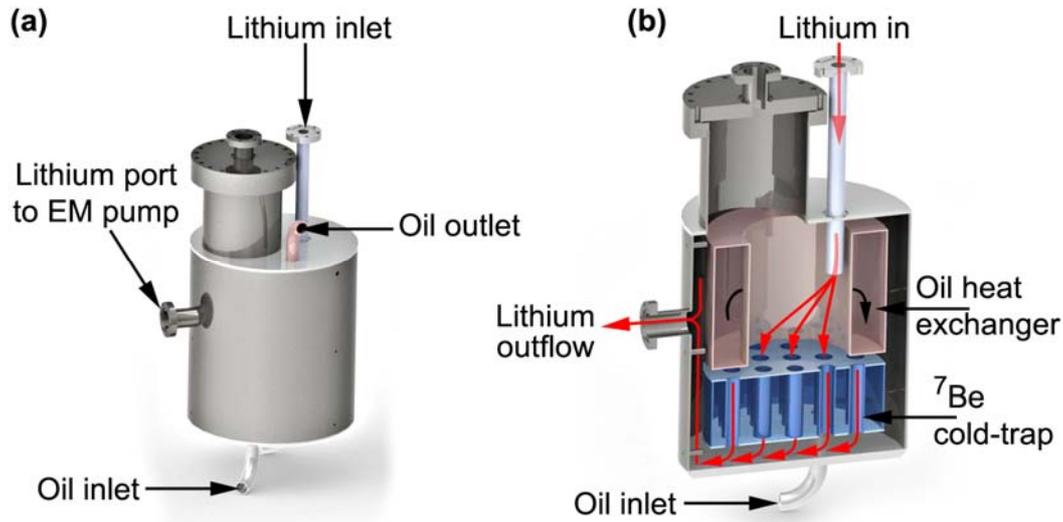

Figure 9: LiLiT lithium reservoir (a) and its cross section (b), depicting the heat exchanger and the $^7$Be cold trap. The liquid-Li flow from the top inlet is indicated by red arrows. The oil circulation from the bottom inlet fills the 7Be cold trap volume and the cooling sleeve as heat exchanger.

## B. The electromagnetic pump

Owing to the high electrical conductivity of liquid lithium, it can be efficiently circulated using externally applied electromagnetic fields. The liquid lithium flow through the circulation loop is driven by a custom-made electromagnetic (EM) induction pump (Fig. 10) consisting of rotating permanent magnets. The liquid lithium passes through a thin loop (rectangular cross section, Fig. 10, A), which is placed between the two magnetic rotors (Fig. 10, B), each of which includes three pairs of permanent magnets of alternating polarity (Fig. 10, C). The alternating magnetic field induces a force that circulates the lithium through the loop. The direction of the magnets rotation is the same as the liquid lithium flow. The frequency of the magnet rotation is between 500 - 1500 rpm, corresponding to lithium velocity of 3 - 7 m/s through the nozzle (see next section).

## C. Flow meter

The flow-meter, mounted above the EM pump, is composed of a permanent magnet and two conductor electrodes attached to the lithium tube, perpendicular to the direction of the magnetic field (Fig. 8, H). The liquid-lithium flow velocity can be estimated from the measurements of the induced voltage generated between the conductors in the constant magnetic field. An average magnetic field of 0.34 T was measured between the permanent magnet poles. During the off-line tests described below, voltages of 1.6 - 3.8 mV were measured, corresponding to estimated lithium flow velocities of 0.4 – 0.6 m/s in the pipe and ~ 3 - 7 m/s in the nozzle.

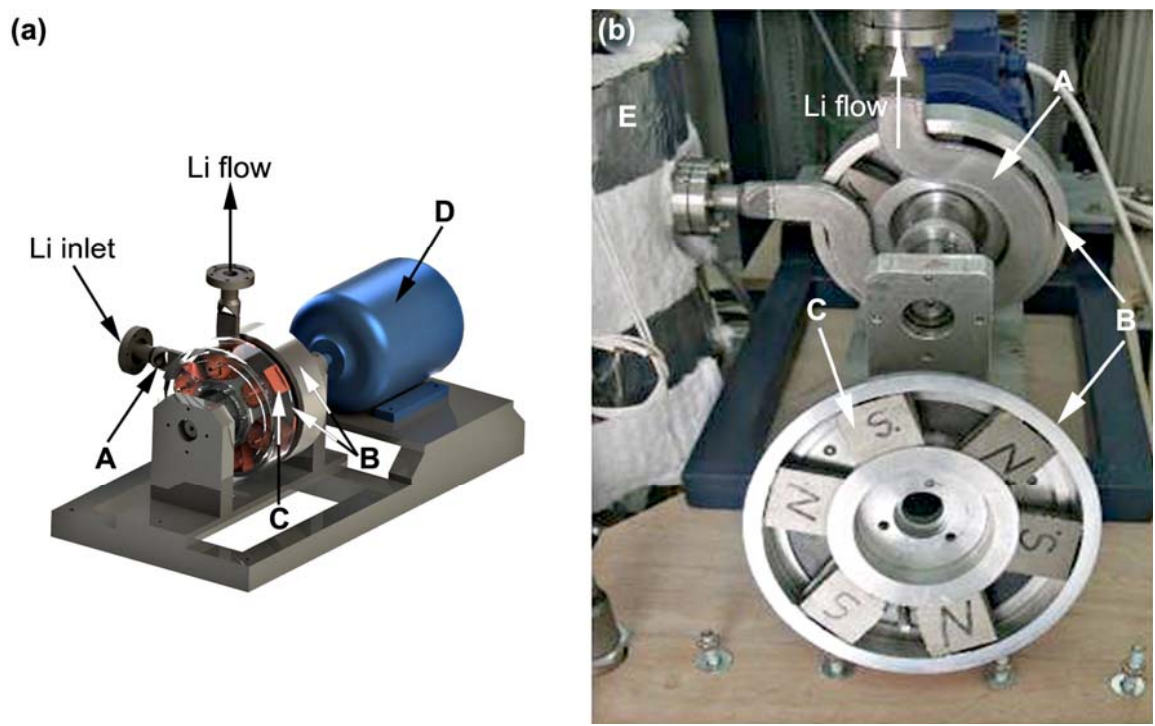

Figure 10: Schematic drawing of the LiLiT electromagnetic pump (a) and its photograph as installed in LiLiT, with one of the magnetic rotors taken apart (b): (A) - lithium EM pump loop, (B) - magnet cases, (C) - permanent magnets in a rotor, (D) - rotating electric motor (shown only in the scheme) and (E) - LiLiT reservoir (seen only in the photograph).

## D. Nozzle and target chamber

The lithium tube enters the top port of the target vacuum chamber (Fig. 11(a)). The target chamber includes a proton beam port with a set of 7 rings (4 cm in diameter) designed to trap lithium vapors during target irradiation, two view ports aimed toward the nozzle and two linear feedthroughs (one for a tantalum foil for beam imaging and tuning and the other for a tungsten wire for beam scanning). The nozzle transforms the circular cross section of the flow (~2.54 cm in diameter) into a 1.5 mm thick and 18 mm wide film (see Figs. 1 and 11(b)). The film flows on a back wall concave towards the beam, with a curvature radius of 30 mm (Fig. 11). Buildup of centrifugal pressure [16] in liquid lithium due to the flow curvature (estimated in our conditions in the $10^{-2}$ mbar range at the depth of the proton Bragg peak for flow velocities of 3-7 m/s) is expected to help further reducing the risk of boiling (see Sect. II C). The neutrons exit through the back wall (Fig. 11(D)) which is made of a thin (0.3 mm) stainless steel sheet. Two metal slabs ("ears" in Fig. 11(b)) are welded on both sides of the nozzle, to serve both as beam diagnostics via temperature measurements and beam shield for the outlet flange. Four thermocouples are connected to the diagnostic slabs for temperature monitoring. The slabs' temperature measurements also assist in centering the high-intensity beam on the lithium jet. The nozzle was designed following a series of water experimental simulations, since water at 20 °C has a kinematic viscosity of $1.01 \times 10^{-6}$ m$^2$/s, which is almost equal to that of lithium at 225 °C ($1.04 \times 10^{-6}$ m$^2$/s). Hence, at these temperatures and similar flow geometries, water flow and lithium flow have almost identical Reynolds numbers. In water simulation experiments, the water velocity was up to 26 m/s and the simulation results showed a stable water film with a smooth surface. Water simulations were also made with a nozzle system made of perspex and scaled in dimensions by a factor 6 in order to reproduce both the Reynolds number and the Weber number of the lithium flow (which takes into account of the liquid surface tension). In these conditions, a water flow of 1 m/s, analogous to a Li flow of ~ 7 m/s, was smooth and stable. A wall made of stainless steel foil, 0.5 mm thick and 19 cm in diameter, shaped as a spherical cap, is located ~1 mm beyond the nozzle and seals the target vacuum chamber neutron exit port (figs. 8 and 11(c)). The foil curvature (convex toward the chamber) is in the opposite direction to the nozzle back wall curvature, allowing positioning of samples behind the curved wall and very close to the neutron source, which is important for activation measurements.

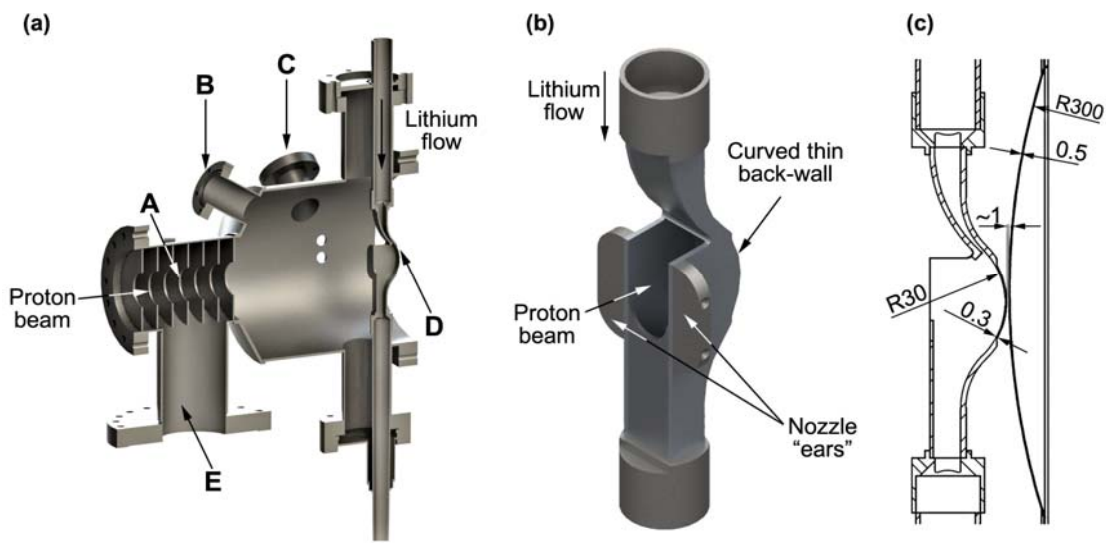

Figure 11: (a) LiLiT vacuum chamber cross section: (A) - set of 7 rings acting as lithium vapor trap, (B) - view port, (C) - port for beam diagnostics, (D) - lithium nozzle, (E) - arc-pump port; (b) and (c) lithium nozzle drawings (dimensions in mm).

### E. Arc pump

The custom-built arc pump (see arc pump port, Fig. 11) is based on sorption of active gas molecules by a layer of Ti getter. Such a pumping method, without moving or fragile parts, is relatively resistant to the lithium vapors that are expected to be present in the target vacuum chamber during liquid lithium irradiation. The pump operated properly during lithium circulation tests and during the electron gun irradiation experiments (see Sec. IV). An ion-pump was attached to the bottom of the arc-pump for pumping argon gas, which cannot be pumped by the arc-pump sorption.

### F. $^7$Be trapping and shielding

$^7$Be radionuclides will be produced in the lithium target through the $^7$Li$(p,n)^7$Be reaction. This nuclide has a half-life of 53 days and emits 478 keV (10.4%) gamma rays. The $^7$Be γ activity produced during continuous irradiation with 1 mA protons at 1.91 MeV is ~ 0.8 γ-mCi/day, reaching saturation of ~ 60 γ-mCi after ~ 7 months of continuous operation. $^7$Be atoms are expected to accumulate in the colder parts of the loop [27], mainly in the cold trap located at the bottom of the reservoir (Fig. 9(b)). A shielding (2 cm thick lead) was designed to reduce the $^7$Be gamma dose and will be placed around the reservoir. With a shielded reservoir the saturation dose rate expected in the working area (30 cm from the system), assuming irradiation of 96 hours/week with a nominal beam (1 mA, 1.91 MeV protons), is lower than 20 μSv/h. We plan to map the $^7$Be γ-activity along the loop and lithium reservoir with the first proton irradiations of the target above neutron threshold.

### G. Control system

The control system of the apparatus is based on a National Instruments compact Field Point (cFP) 2020 controller [28] operated via a PC. The controller governs the following input output (I/O) interface modules:

1. 16 analog current inputs (AI) monitoring the temperatures values at various positions in the LiLiT reservoir, loop and nozzle slabs.
2. Two AI modules (8 inputs each), of analog voltage and current inputs for monitoring the vacuum, lithium velocity (flow-meter voltage), oil velocity, lithium loop and oil loop temperatures at various locations, position sensors and current of the tungsten wire and tantalum diagnostic plate.
3. 8 analog input/output (AI/O) modules for monitoring the lithium and the oil pumps, operating the nozzle direct current heating and the electron gun.
4. 22 digital outputs (DO) for control of lithium and oil heaters, lithium and oil pumps and oil-air heat-exchanger (fans).
5. 25 digital input signals (DI) for monitoring lithium temperature status and controlling the set points of the machine safety system parameters (lithium flow, vacuum, lithium temperature, EM pump operation and scanning wire position).

A graphical user interface (GUI), allows operation and control of the system, data acquisition and recording via a LabView version 8.2 [28] program. Prior to installation of the LiLiT for on-line tests at the SARAF beam line, the controller will be upgraded to a NI compact RIO (cRIO) controller, which includes Field-Programmable Gate Array (FPGA) modules. The new controller will improve performance and reliability.

## IV. OFF-LINE EXPERIMENTS WITH THE LiLiT SYSTEM
### A. Circulation tests

A stand-alone fire resistant offline laboratory was built to develop proper working and operation procedures, perform preliminary experiments with the liquid-lithium apparatus and establish the safety regulations for future handling of the online system. 7.5 kg of lithium (99.9% min.) were loaded into the lithium tank (Fig. 9) under a controlled dry (relative humidity less than 5%) argon atmosphere and melted at ~220°C. Several failures occurred during the first trials of liquid lithium circulation. In the first circulation experiment, the EM pump rotating frequency was increased gradually up to ~500 rpm. At this stage lithium was pushed through the tubes, reached the vacuum chamber but immediately solidified at the nozzle edge, apparently due to the lower temperature of the nozzle inside the vacuum chamber. A heating system (~1 kW) was developed to preheat resistively the nozzle up to a temperature to ~ 200°C by connecting high-current leads from a current transformer directly on top and bottom ends of the nozzle tubing outside the vacuum chamber. A similar resistive heating was applied at both ends of the EMP loop in order to melt the solid lithium that fills the loop before operation. At this stage, the resistive heating in the loop area is complemented by eddy-current heating by operating the EMP in reverse direction. In a following experiment, the lithium flowed through the nozzle but immediately surged back upward into the vacuum chamber. The reason for the overflow appeared to be incomplete lithium liquefaction in the reservoir, causing a clog in the reservoir lithium inlet. From these experiments it was concluded that for proper circulation, all parts of the loop, especially the nozzle, must be carefully preheated to a temperature above 200°C for a few hours to ensure complete liquefaction of the lithium in the reservoir. In a third attempt, under the conditions described above, we were able to demonstrate stable circulation. The lithium flow is started at a velocity above 1 m/s by operating the EMP at the appropriate speed and creates immediately a lithium film adhering to the supporting back wall of the nozzle under stable vacuum conditions (~$10^{-6}$ mbar, see fig. 14 for location of vacuum gauge), lasting for long periods (~10 hours). When circulating liquid lithium for the first time after the system had been filled with Ar (*e.g.* for servicing a component), we observed short bursts of pressure (up to $10^{-4}$ mbar), attributed to Ar outgassing. The Li film was observed through a view port (fig. 12, see also [29] for the movie of a recent circulation test). No splashes of lithium droplets or aerosols formation were observed up to the maximum flow velocity we operated (~ 7 m/s). At low flow velocity, the lithium film showed slight waviness (considered insignificant relative to the film thickness) but with the increase of EM pump power and film velocity, it became smoother and more stable. Our present estimate of the largest flow velocity attained is ~7 m/s in the nozzle section, based on a measurement using the electromagnetic flow-meter (Sec. III C).

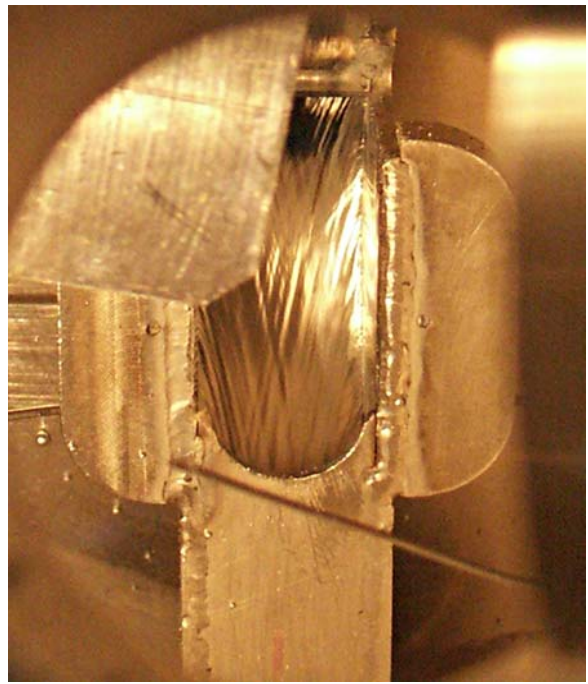

Figure 12: The 18 mm wide lithium film, as photographed through the vacuum chamber view port, flows on the concave supporting wall of the lithium nozzle. The flow velocity is estimated as 2.5 m/s based on the flowmeter reading. The tantalum plate used for beam positioning is seen near the upper left corner of the lithium nozzle. When extracted, the plate clears the beam path completely. See also [29].

### B. Electron-gun tests

The second aim of the off-line tests at the LiLiT fire resistant laboratory was to demonstrate that an electron beam power equivalent to a 1.91 MeV, > 2 kW proton beam (see Sect. II, figs. 4 and 5 and table I) can be dissipated by the circulating lithium.

#### 1. Electron beam similarity to proton beam

A high-intensity (26 kV, 2.6 kW) electron gun was built to create an electron beam capable of simulating the energy deposition of a 1.91 MeV, >2 kW proton beam. The two-dimensional profile of the power deposition of 26-keV electrons in liquid lithium with transverse distribution close to a radial Gaussian and standard deviation σ = 2.7 mm, was calculated using the Monte Carlo simulation code CASINO [30]. The electron beam power density distribution for a 1.56 kW (60 mA) electron beam is shown in fig. 13 with a peak power density around ~2 MW/cm$^3$, 25 μm within the liquid lithium. The electron power distribution (fig. 13) is very close to that of the nominal proton beam power shown in fig. 5 in the Bragg peak area. The calculated peak of power density for the electrons, ~2 MW/cm$^3$ is equivalent to that of a radial Gaussian proton beam with standard deviation σ = 2.8 mm, energy of 1.91 MeV and intensity ~3 mA (~0.67 MW/cm$^3$/mA). The maximum energy deposition density is one of the most significant parameters governing lithium jet disturbances. We anticipate that the behavior of our lithium film, its stability, continuity and shape, under such a high power density electron beam, provides a reliable test for its behavior under the proton beam of SARAF.

2. **Electron gun experiments**

The high-intensity electron gun was attached to the LiLiT system together with a solenoidal coil for electron beam focusing and a set of 4 coils placed symmetrically around the beam axis for beam deflection (Fig. 14). The coils were located about 45 cm upstream of the nozzle. In a typical experiment, a low-power electron beam (around 10 W) was first centered and focused on the tantalum foil block (Sect. III D), which was moved in front of the lithium nozzle. The beam profile was then measured using the tungsten wire scanner. Following beam tuning, the electron gun was shut down and the tantalum foil was removed from the beam. After the lithium flow was established and the lithium film on the nozzle back wall was confirmed visually, the electron gun was turned on again. In that phase video recordings were made of the beam-on-target image, while increasing the lithium velocity and the electron beam intensity. Above ~200 W, an orange glow (possibly from reflection of glow of the electron-gun filament) was observed on the liquid lithium, when the chamber lights were off (Fig. 15(b)). No other glow, disturbances or changes in the lithium jet were noticed during the experiments. In the first experiment the electron beam power on the flowing lithium film was increased up to 2.2 kW (85 mA). The electron beam shape (measured at low intensity by the scanning wire) could be fitted to a Gaussian with sigma of 3.5 - 4 mm. This implies a maximum power density of 2.85 kW/cm$^2$ applied to the liquid lithium, flowing at an estimated velocity of 3.2 m/s. However, throughout this first experiment the system was not stabilized and the irradiation periods at high power were short, limited by continuous heating of the nozzle diagnostic slabs by the electron beam halo (up to temperatures above 700°C without reaching equilibrium) and target pressure increase above $2\times10^{-5}$ mbar (set as the maximum pressure allowed). Another problem, possibly due to the overheating of the nozzle walls by the beam halo, was excessive lithium evaporation at high electron beam power, covering the internal parts of the target chamber and blocking the view ports. The temperature of the lithium bulk was measured in the reservoir in 9 different locations. During this experiment, the oil was circulating through the reservoir but the oil-air heat exchanger was not operating.

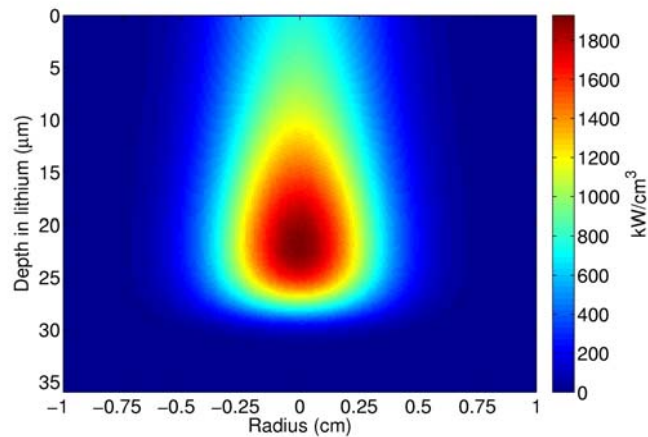

Figure 13: Two-dimensional density plot of power deposited in lithium by a 26 keV, 60 mA electron beam, with transversal distribution close to radial Gaussian and standard deviation σ = 2.7 mm. The power distribution was calculated with the Monte Carlo CASINO simulation code [30].

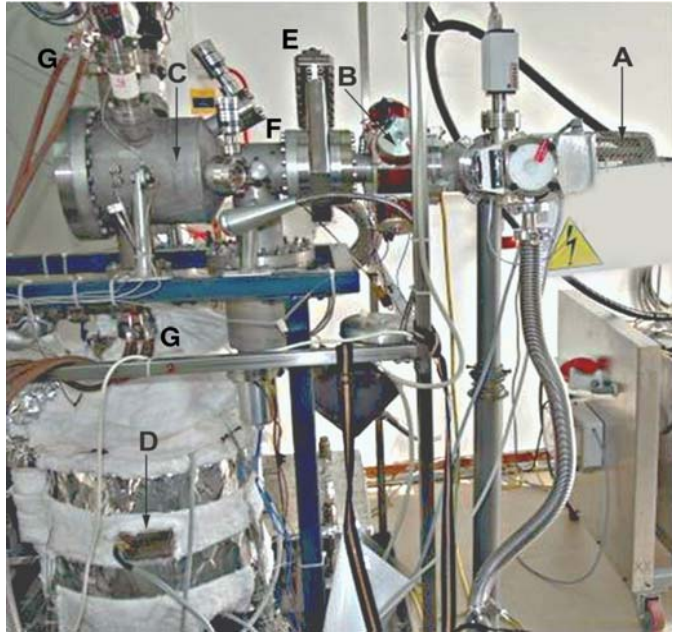

Figure 14: Electron gun attached to the LiLiT system: (A) - electron gun, (B) - focusing/deflection coils; (C) - target vacuum chamber; (D) - lithium reservoir covered by the heater blankets; (E) - gate valve; (F) – vacuum gauge; the gauge (not seen in photograph) is located behind the nipple connecting vacuum chamber and gate valve at a distance of ~ 25 cm from the nozzle; (G) - high-current leads for resistive pre-heating of the nozzle.

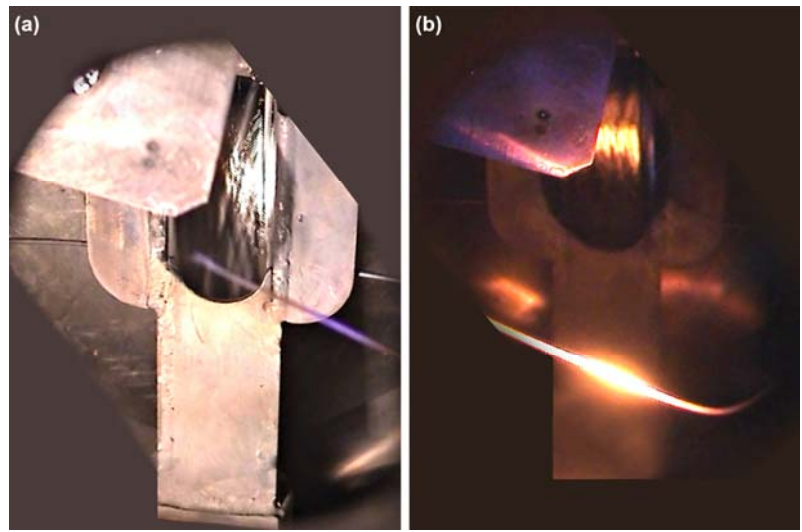

Figure 15: Photographs of the lithium flow through the nozzle during electron gun irradiation, with target chamber lights on (a) and off (b) taken from the chamber viewport (see text). The pictures were taken from a diagonal angle since the e-gun was attached to the front view port. The tantalum diagnostic plate (upper left) is covering part of the viewing angle of the nozzle but clears the electron beam.

In order to overcome the overheating of the nozzle walls by the electron beam halo and excessive lithium evaporation, the focusing and deflection coils were upgraded, producing a well-centered near-Gaussian electron beam profile with σ ~ 2.7 mm. A cold trap between the target chamber and the view ports was added in order to adsorb lithium vapors and avoid obscuring of the view ports. During the second experiment, four irradiations (more than 10 minutes each) were performed (see fig. 16, irradiations are marked 1-4). The electron beam power was increased from an average of 0.7 kW in the first run (fig. 16, # 1) to an average of 1.5 kW in the fourth run (fig. 16, # 4), fluctuating between 1.4 and 1.9 kW due to instability of the e-gun power. During these irradiations the temperatures of the nozzle diagnostic slabs were stable at values up to 350°C and 450°C at the coldest and warmest measurement points on the slabs (fig. 16). The vacuum in the target chamber was about $1\times10^{-5}$ mbar and even improved during the last irradiation to $6\times10^{-6}$ mbar (fig. 17). No disturbances were observed in the lithium flow. The system was voluntarily stopped after 45 minutes and no indication for excessive lithium evaporation was found. The average areal power density applied to the liquid lithium, flowing in the nozzle at an estimated velocity of ~4 m/s, was 4.2 kW/cm$^2$ and the average volumetric power density (for 60 mA, see fig. 13) is estimated to be ~2 MW/cm$^3$.

The oil-air heat exchanger was not operated during these irradiations either, although the oil circulated through the lithium reservoir. The minimum and the maximum bulk lithium temperatures in the reservoir are shown in Fig. 17. The temperature in the colder (lower) areas of the reservoir increased by 10°C, from 190°C to 200°C. The warmer (upper) areas of the reservoir (with temperatures above ~220°C) were not affected by the electron irradiation (Fig. 17).

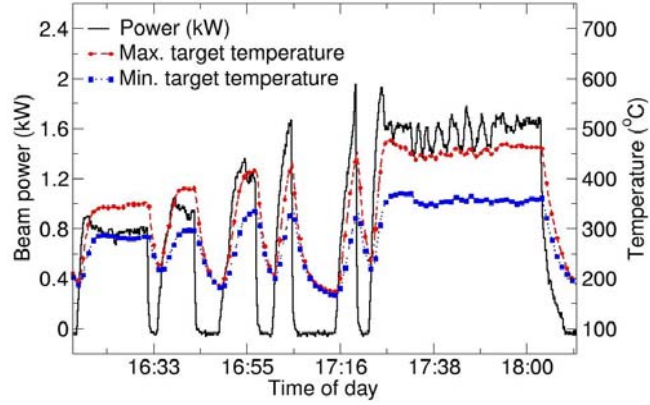

Figure 16: Electron beam power and maximum and minimum temperatures measured on lithium nozzle diagnostic slabs during the final electron beam experiment (see text).

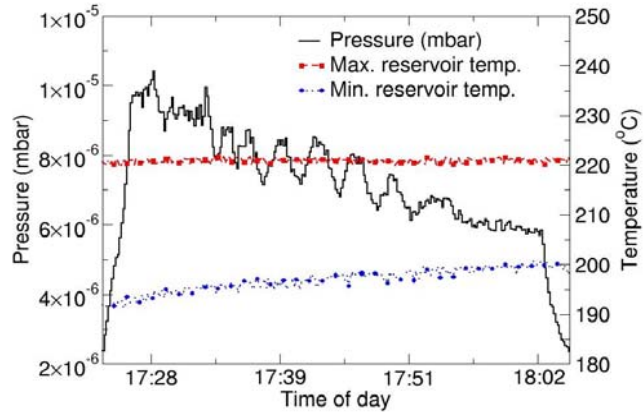

Figure 17: Vacuum in the LiLiT target chamber (see fig. 14 for location of vacuum gauge) along with the maximum and minimum lithium temperatures in the lithium reservoir during the final electron beam irradiation (see text).

## V.  SIMULATION OF NEUTRON YIELD, ENERGY SPECTRUM AND ANGULAR DISTRIBUTION FROM THE LiLiT SETUP

The LiLiT setup is presently being installed at the SARAF linac toward commissioning experiments with the high-power proton beam of SARAF. Detailed neutron transport simulations, taking into account the realistic geometry of LiLiT and its surrounding in the SARAF accelerator hall, were performed in order to estimate the neutron yield, energy spectrum and angular distribution expected at a secondary target located 3 mm downstream the LiLiT neutron port (Fig. 8(C)). The calculations were made using methods described in [6] for a 1.91 MeV proton beam with energy spread of 15 keV. The results are presented in Fig. 18 and show that a neutron intensity of $2.4 \times 10^{10}$ n/s is achievable with a 1 mA proton beam from SARAF, more than one order of magnitude higher than similar available sources, with a most probable and mean energy of 28 keV and 46 keV, respectively.

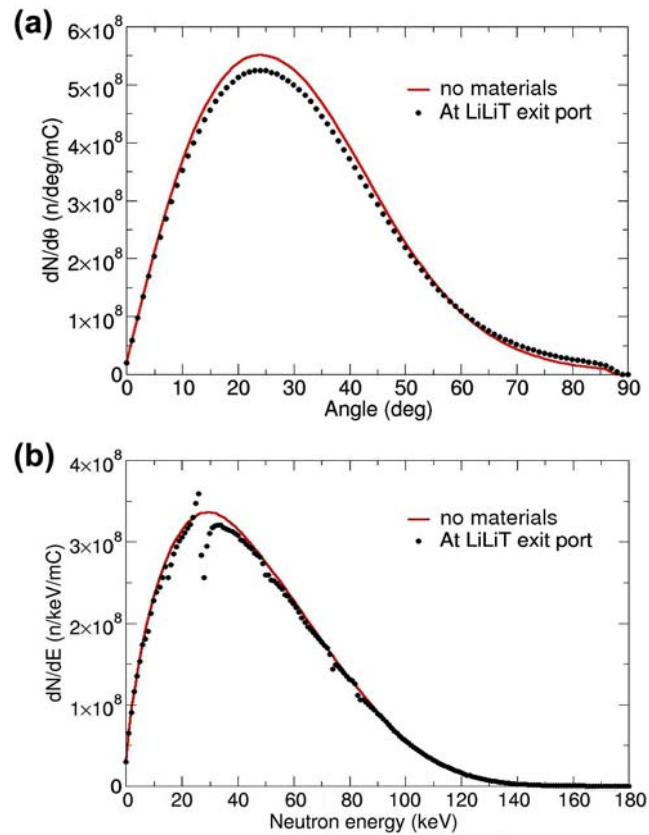

Figure 18: Simulations of the neutron angular distribution (a) and energy spectrum (b) integrated over all angles as seen by a secondary target behind the LiLiT exit port, 3 mm downstream the lithium surface (solid dots). The simulations include the effect of structural and surrounding materials. The distributions are calculated for a proton beam ($E_p$ = 1910 ±15 keV (black dots) as available at the SARAF accelerator (see text). The solid lines represent the distributions of the $^7$Li(p,n) yield with no material interactions expected at the same location. The effect of the $^{56}$Fe resonance from structural materials can be seen at ~27 keV. The total neutrons rate incident on the secondary target is $2.4 \times 10^{10}$ n/mA/s.

**SUMMARY**

The development and off-line experiments of a liquid-lithium target (LiLiT) for high-intensity proton beams at energies above the (p,n) reaction threshold, are described. The challenge in such a design is the large volume power densities (>1

MW/cm$^3$) created by a narrow Gaussian beams (σ < 3 mm), required for the desired epithermal neutron flux density.

The system was designed based on a thermal model that estimated the required jet velocity needed to prevent excessive evaporation and boiling of the liquid lithium. Lithium circulation experiments confirmed that the system can operate safely and that at high velocity (~7 m/s) the forced liquid-lithium flow on the concave supporting wall circulates in stable conditions in a vacuum environment. An electron gun was used to apply high power densities on the target. With a continuous electron beam power of 1.5 kW irradiating the target, the liquid lithium target was shown to dissipate areal power densities > 4 kW/cm$^2$ and a volume power density ~ 2 MW/cm$^3$ at a lithium flow velocity of ~ 4 m/s, while maintaining stable temperature and vacuum conditions. Based on calculated temperature profiles, we anticipate that for ion beams (where the Bragg peak is located deeper inside the lithium flow) surface temperatures, while remaining a limiting factor, will still allow continuous operation of the target.

High-intensity narrow proton beam irradiation (σ = 2.8 mm, 1.91-2.5 MeV, 2 mA) of LiLiT is under commissioning at the SARAF superconducting linear accelerator. Simulations show that a neutron intensity of $2.4 \times 10^{10}$ n/s is achievable with a 1 mA proton beam of ~1.91 MeV from the SARAF accelerator. The setup at SARAF will be used for nuclear astrophysics research and, together with a beam shaping assembly, to demonstrate the applicability of this new concept to accelerator-based BNCT. We plan to use LiLiT in the future also with a deuteron beam for fast neutron production and radioactive ion production. The experience gained with LiLiT will serve in the design of an upgraded target matching the higher energies (40 MeV deuterons) of the final configuration of SARAF (Phase II).

## ACKNOWLEDGMENTS

We are grateful to Y. Momozaki, J. Nolen and C. Reed from Argonne National Laboratory for the help extended to us, to W. Gelbart for his early contributions and to R. Reifarth for his collaboration through the German-Israeli Foundation (GIF). We gratefully acknowledge the support of the Pazi Foundation and GIF.